# Mechanoluminescence in crystalline inorganic materials: local disorder and the elastic distortion hypothesis

T. Rouxel[1,2], X. Rocquefelte[3] and S. Tanabe[1]

[1]Graduate School of Human and Environmental Studies, Kyoto University, Yoshida-nihonmatsu-cho, Sakyo-ku, Kyoto 606-8316, Japan.

[2]Glass Mechanics, Physics Institute, UMR-CNRS 6251, Beaulieu campus, University of Rennes, 35042 Rennes cedex, France.

[3]CNRS, ISCR (Institut des Sciences Chimiques de Rennes) - UMR 6226, F 35000 Rennes, France.

**Abstract –** In this exploratory work, we aim to understand the mechanoluminescence (*ML*) phenomenon manifested by various inorganic compounds by focusing on the influence of mechanical loading on local distortion and loss of symmetry at active sites. To this end, we have analyzed the elastic deformation of several relevant crystalline phases and shown, through a kinematic analysis based on elastic constants, that the loading-induced distortion is smaller than, but remains comparable in magnitude to, the intrinsic structural distortion as quantified by the Baur descriptor. Although the structural distortion has previously been proposed as a structural fingerprint of the *ML* potential of a compound, it is by nature a static parameter, and must be supplemented by a dynamic distortion contribution that develops under mechanical load. By approximating the crystal deformation by that of simple polygonal motifs, we have been able to propose a rationale for otherwise puzzling observations: in particular, the marked differences in sensitivity to hydrostatic pressure and to shear, the appearance of an intense *ML* peak upon unloading in certain compounds, and the contrast in behavior depending on whether the preliminary *UV* irradiation is performed under load or at rest.

## I. Introduction

Mechanoluminescence is the ability of materials to emit light under mechanical loading. It can be observed during friction (triboluminescence), at fracture (fractoluminescence), or under ordinary mechanical loading and chiefly reversible experiments in compression, tension, shear and hydrostatic pressure (elastico-mechanoluminescence, *EML*). It generally requires *UV* light irradiation prior to testing. This phenomenon may in fact be quite universal, but is often too weak to be visible to the naked eye. Most of the studies carried out to date focused on the physics of the electron transitions, the dynamics of the point defects and of the trapping-detrapping processes, and the dependence of all these atomic-scale mechanisms on the crystal structure and chemistry of the compound. Although recent studies conclude to the importance of the crystal symmetry and local disorder, there are no report dealing with crystal distortion and local disorder changes that may result from the macroscopic stress field. However, these changes are likely to be responsible for the notable differences observed between different crystal structures for a given compound chemistry (stoichiometry), and there are obvious relationships between crystal distortion, crystal field strength, and the applied stress field. These relationships are discussed in the present paper, independently of the fine details of the atomistic mechanisms, tunnelling effects, electric polarization or trap dynamics, which are consequences of the stress field and have already been described in previous review papers on *ML* mechanisms [1-6]. The aim of this article is to provide plausible explanations to phenomena that have been widely observed but remain unelucidated, such as light emission under both loading and unloading, the disappearance of *EML* under pure shear or pure hydrostatic pressure, and the changes observed when *UV* irradiation is applied under mechanical loading, to list a few of them. There are actually very few reported attempts to correlate *EML* to the macroscale stress-field ($\tilde{\sigma}$), and with the distinct contributions of pressure (hydrostatic part of $\tilde{\sigma}$) and shear (deviatoric part of $\tilde{\sigma}$). In this paper, by comparing a series of chemical systems exhibiting *ML* in which such observations were made, and by examining the kinematics of crystal elastic deformation, we aimt to shed light on possible mechanisms driving *EML* and to provide some thoughts and directions for future investigations.

## II. Ten key observations

### II.1 Mechanical testing experiments

In what follows, ten key observations, referred to as ***KO1***,…***KO10*** for convenience, were identified from a literature review on the occurrence of *ML*, and will be discussed in § III and IV. Some of them are illustrated in Fig. 1. First, in most cases the material must be irradiated with *UV* light before the phenomenon can be observed (***KO1***); Second, the signal emitted under load is generally much weaker than that of persistent luminescence (***KO2***), which necessitates waiting before applying the load so that the *EML* intensity is not masked by the persistent luminescence (*PL*) signal. Nevertheless, it is noteworthy that there is no one-to-one relationship between *PL* and *EML* performances, and several compounds exhibit *PL* without any detectable *ML*. For example, $Y_2O_3:Eu^{3+}$ and $Y_2O_2S:Eu^{3+}$ phases belong to the same crystal system and both exhibit *PL*, but *EML* is significant in the latter and absent in the former [7]. A luminescent center, such as $Eu^{2+}$ in $SrAl_2O_4$, is generally required, but a co-dopant such as $Dy^{3+}$ in $SrAl_2O_4$ typically very beneficial to *ML* (***KO3***). Depending on the material, an *EML* peak may also appear during unloading (***KO4***), as in $ZrO_2:Ti^{4+}$ [8], $Sr_3Al_2O_6:(Eu^{2+}, Dy^{3+})$ [8], $ZnS:Mn^{2+}$ [9][Ali, 2015], $BaSi_2O_2N_2:Eu^{2+}$ [10], and $SrAl_2O_4:(Eu^{2+}, Dy^{3+})$ [11] (Figs 1a-c). The emitted intensity is most often proportional to the mechanical power developed during the test (***KO5***) [3, 11- 13]. In other words, the loading must be sufficiently rapid for the signal to be intense, which is why ball-droping and scratching experiments are usually effective at revealing mechanoluminescence. Nevertheless, other, lesser-known characteristics have also attracted our attention because they are directly linked to a fundamental understanding of how a mechanical field affects a material. These are the seemingly independent, but very likely connected (see § II.3), influences of the hydrostatic (pressure) and deviatoric (shear) components of the stress field (***KO6***). Crystals belonging to the same crystal system (for example, monoclinic $SrAl_2O_4$ and $Ba_4Si_6O_{16}$, illustrated in Fig. 1) can thus be more sensitive to pressure than to shear, or vice versa. For instance, no *EML* signal is observed under pure shear in $Ba_4Si_6O_{16}$, whereas this compound exhibits a relatively strong *EML* response under hydrostatic pressure of the same order (less than 100 MPa) [13, 14]. On the contrary, the signal under hydrostatic pressure is weak in $SrAl_2O_4$ compared with that observed under uniaxial compression (pressure + shear) or torsion (shear) [12, 15]. Furthermore, $Ba_4Si_6O_{16}$ shows no intensity regain during discharge (Fig. 1d), and a decrease in intensity is observed during decompression under hydrostatic load for both $Ba_4Si_6O_{16}$ and $SrAl_2O_4$ (Figs. 1e and 1f). Besides, irradiation under mechanical load might lead to a different *EML* response (***KO7***), such as an intense peak upon unloading in uniaxial compression and a weakening of the *EML* intensity upon re-loading (Fig. 1g) [12], or the absence of a peak upon unloading under hydrostatic pressure (Fig. 1h). Notably, the minimum intensity during unloading (in $Ba_4Si_6O_{16}$:($Eu^{2+}$, $Ho^{3+}$)), or during reloading of a sample irradiated under load ($SrAl_2O_4$:($Eu^{2+}$, $Ho^{3+}$) does not occur at the load extremum - i.e. at complete unloading or reloading -, but at an intermediate load (about half the unloading step in the case of $Ba_4Si_6O_{16}$) (***KO8***) [14], as illustrated in Fig. 1d for example. In addition, a threshold stress for the onset of *EML* has also been reported in some cases (***KO9***), of the order of 3 MPa for ZnS:$Mn^{2+}$ under hydrostatic pressure [16]. Chandra et al. [16] linked this observation to the generation of a piezoelectric field, which is often invoked to explain mechanoluminescence [3, 17] and would correspond to the critical stress required to generate an internal electric field strong enough to release trapped charge carriers, or to produce sufficient nonlinear lattice deformation to induce dislocation activity or internal friction. Finally, local symmetry breaking at active sites appears favorable to *ML* (***KO10***) [3, 7, 18].

Taken together, these observations show that the underlying mechanisms go beyond the classical trap-detrapping via piezoelectric field, defect/dislocation-mediated *ML*, phase transitions, and interface effects in composites. At the first place, this brings to light the paramount importance of the crystal response to both pressure and shear, in terms of volume change and distortion, and of how the mechanical field is transferred to the active sites (luminescent centers and traps). This is where kinematics comes into play, as discussed in § IV.

### II.2 Tendencies through different chemical systems

The systems described below and illustrated in Fig. 2, within which mechanoluminescent phases were obtained, are not exhaustive. The selection was guided, on the one hand, by the authors' experience in the field of *ML* materials and, on the other hand, by the number of published studies, which bring to light some important trends. In most cases, a rare-earth cation, such as $Eu^{2+}$, is responsible for the fluorescence, phosphorescence (persistent luminescence) and mechanoluminescence as well. It is most likely that $Eu^{2+}$ replaces $Sr^{2+}$, and as europium is introduced in the 3+ valence state ($Eu_2O_3$), charge neutrality is maintained by forming Sr vacancies, $V_{Sr}''$ while oxygen enters normal oxygen lattice sites ($O_O^x$). The following reaction takes place:

$$Eu_2O_3 \rightarrow 2Eu_{Sr}^{\bullet} + V_{Sr}'' + 3O_O^x \qquad (1)$$

where, $Eu_{Sr}^{\bullet}$ is a $Eu^{2+}$ cation on a Sr site. Another possibility is that, because the synthesis atmosphere is typically reducing (e.g., an $N_2$-$H_2$ mixture), EuO forms during processing, since $Eu^{2+}$ substitutes for $Sr^{2+}$, no additional point defect is required for charge balance. In the electron trapping/detrapping mechanism, which is most often invoked to explain both *PL* and *ML*, pre-irradiation (typically with *UV* light, whose energy exceeds the band gap) is generally required prior to mechanical testing to observe *ML*. The reason for this, answering ***KO1***, is that the population of electrons trapped in the activatable deep *ML* traps is small at the equilibrium under ambient conditions. Furthermore, the density of traps deep enough to retain electrons against thermal activation is much smaller than that of the shallow traps responsible for *PL*, which accounts for ***KO2***. Although not discussed in detail in this paper, the *ML* intensity is most often enhanced by the incorporation of a co-dopant. For example, with $Eu^{2+}$ as the activator and $Dy^{3+}$ as a co-dopant, the *ML* intensity of $SrAl_2O_4$, $BaAl_2O_4$, and $CaMgSi_2O_7$ is reported to be about two orders of magnitude higher than in the Dy-free counterparts. This enhancement is usually attributed to the creation of shallower traps (relative to those associated with $Eu^{2+}$ alone), from which carriers are more easily released under mechanical loading. The formation of charge-compensating point defects associated with $Dy^{3+}$ incorporation further increases the trap density. Together, these effects offer a tentative explanation for ***KO3***.

The lattice distortion, especially at the cationic sites of interest, is often considered to play a key role in *ML*. The structural distortion index ($D_s$) introduced by Baur [19] has been adopted by several authors to describe the degree of distortion of the coordination polyhedron around the luminescent center in *ML* materials [7, 18, 20]. For a given polyhedron, $D_s$ is calculated from the coordination number (*CN*) of the "central" cation, the bond length between the cation and each coordination anion ($l_i$), and the mean bond length ($\overline{l_\iota}$) as follows:

$$D_s = \frac{1}{CN}\sum_{i=1}^{CN}\frac{\left|l_i - \overline{l_\iota}\right|}{\overline{l_\iota}} \qquad (2)$$

In the present study, crystallographic data extracted from *CIF* (Crystallographic Information File) format files were used to estimate $D_s$. The structure was analyzed to ensure consistent identification of inequivalent sites and coordination by importing the *CIF* file into a Python code (Python 3.9.13) and using Pymatgen. Coordination numbers (*CN*) were determined automatically using CrystalNN for each cationic site of interest, and were compared with existing experimental structural data. Parameter $D_s$ was computed using the set of nearest-neighbor anions determined by CrystalNN. Manual specification of the cutoff distance was used to test the sensitivity of $D_s$ to neighbor selection, and in particular to obtain $D_s$ for a specific (known) cationic coordination. Results for $D_s$ are given in Table 1. The values obtained for the two inequivalent sites of Sr in the monoclinic phase of $SrAl_2O_4$ ($D_s(Sr_1) = 0.065$ and $D_s(Sr_2) = 0.073$) are in agreement with those reported by Denault et al. [20] from *XRD* measurements at room temperature (0.0465 and 0.0734 respectively for $Sr_1$ and $Sr_2$), and so are the values for BaZnOS et CaZnOS (0.063 and 0.148 for Ca and Ba respectively) with the experimental ones by Lin et al [Lin, 2022] (0.062 and 0.141). The results for the low temperature phase of $BaAl_2O_4$ ($D_s(Ba_1) = 0.058$ and $D_s(Ba_2) = 0.027$) are also close to the experimental data from Denault et al. in the case of ($D_s(Ba_1) = 0.058$ and $D_s(Ba_2) = 0.039$). It is noteworthy that $D_s$ calculation is sometimes questionable as the coordination number might be ambiguous. This happens for example in $SrAl_2O_4$, as Sr might be considered as 7, 8 or 9-fold coordinated, or in barium silicates, where Ba is mostly 8-fold coordinated but some sites can be described as 6-fold coordinated. Nevertheless, by limiting the precision on $D_s$ to three decimal places, these results remain relevant for comparing compounds from different chemical systems and for analyzing the trends. Let us now discuss the structure of a series of compounds from different chemical systems (Fig. 2) and the *ML* observations that were reported for these compounds.

a) $SrAl_2O_4$:($Eu^{2+}$, $Dy^{3+}$)

Both the hexagonal and monoclinic $SrAl_2O_4$ phases are built on a 3*D* framework of interconnected $AlO_4$ tetrahedra and distorted $SrO_7$ polyhedra. In the hexagonal phase ($P6_3$, high-temperature), the $AlO_4$ tetrahedra form a regular 3*D* framework. They are connected corner-to-corner in a repeating, nearly ideal pattern, leaving regular channels along the c-axis, which host $Sr^{2+}$ ions. The Al–O–Al angles are relatively uniform, and the $Sr^{2+}$ cages are weakly distorted [21]. By contrast, in monoclinic $P2_1$ phase (room-temperature), the $AlO_4$ tetrahedra undergo tilting and rotation, producing a wide distribution of Al–O–Al angles, while $Sr^{2+}$ occupies more irregular cavities. The $P2_1 \rightarrow P6_3$ transition is displacive and driven by the collective tilting (5 to 10°) of the $AlO_4$ tetrahedra [22]. This distortion enhances the crystal field splitting at $Eu^{2+}$, which substitute for $Sr^{2+}$ owing to their nearly identical ionic radii ($r_{Sr2+}$ = 1.32 Å and $r_{Eu2+}$ = 1.31 Å), and contributes to both long persistent luminescence and *ML* [23]. In this compound, *ML* is generally attributed to the detrapping of electrons from oxygen vacancy sites, some of which are particularly sensitive to both shear and pressure [12]. *ML* is observed only in the monoclinic phase, possibly due to its greater disorder, as reflected in the larger values of $D_s$ and its lower symmetry.

b) $BaAl_2O_4$:($Eu^{2+}$, $Dy^{3+}$)

Both the low-temperature (< 400 °C) and high-temperature (> 400 °C) phases are hexagonal, but only the low-temperature $P6_3$ phase (*LT*), which lacks inversion symmetry and give rise to piezoelectricity, exhibits *ML* [3, 24, 25], whereas the high-temperature $P6_322$ phase (*HT*), with its higher symmetry, does not. The basic framework consists of corner-sharing $AlO_4$ tetrahedra forming channels in which $Ba^{2+}$ ions reside. The lower symmetry and larger structural disorder of the *LT* phase may account for its *ML* response. Specifically, the larger distortion of the *LT* phase ($D_s(Ba_1) \approx 0.058$ for *LT* vs. 0.017 for *HT*) creates localized traps and an internal stress field, enhancing the trapping and release of charge carriers [26-28]. When $BaAl_2O_4$ is doped with $Eu^{2+}$, the latter substitutes for $Ba^{2+}$, introducing additional lattice distortion owing to the difference in ionic radii ($r_{Ba2+}$ = 1.42 Å and $r_{Eu2+}$ = 1.31 Å).

c) ($Ca_2$, $Ba_2$)$MgSi_2O_7$:($Eu^{2+}$, $Dy^{3+}$)

Undoped $Ca_2MgSi_2O_7$ crystallizes in the tetragonal melilite structure (space group $P\bar{4}2_1m$) and exhibits *ML* upon co-doping with $Eu^{2+}$ and $Dy^{3+}$, which substitute for $Ca^{2+}$ (or $Ba^{2+}$) and induce lattice distortions because of the substantial difference in ionic radii ($r_{Ba2+}$=1.42 Å and $r_{Ca2+}$=1.12 Å and $r_{Eu2+}$=1.31 Å) [29-32]. When Ba substitutes for Ca, the space group is reduced to C2/c, of lower symmetry. This monoclinic system is not a sub-group of $P\bar{4}2_1m$, but a reconstructive change in the tetrahedral sheet topology [33]. Interestingly, in the melilite family ($A_2MgSi_2O_7$, where *A* = Ba, Sr, or Ca), which consists of layered [$MgO_4$] and [$SiO_4$] tetrahedra arranged in ring-like units, with the $A^{2+}$ cations sitting in large, irregular cavities, the degree of structural distortion and disorder increases as the size of the *A*-site cation decreases. Accordingly, $Ca_2MgSi_2O_7$ is the most distorted ($Ca^{2+}$ being the smallest cation, $Ba^{2+}$ the largest), and $Ba_2MgSi_2O_7$ exhibits the weakest *ML* signal. $Sr_2MgSi_2O_7$, although not shown in Table 1, occupies an intermediate position. The Ca-melilite structure, the most active-*ML,* accommodates the undersized $Ca^{2+}$ ion through pronounced tetrahedral tilting, rotation, and bond-angle distortion. As a matter of fact, $D_s$ is much larger for the Ca- than for the Ba-melilite ($D_s$ = 0.058 and 0.018, respectively)

d) $ZrO_2:Ti^{4+}$

Titania containing zirconia adopts the monoclinic form ($P2_1/c$) at low Ti content (below 0.5 at.%), transforms toward a tetragonal phase ($P4_2/nmc$) with increasing Ti content, and ultimately toward a rutile phase (natural occurrence for pure $TiO_2$) at high content [34-36]. With an ionic radius of ~0.605 Å, $Ti^{4+}$ is smaller than $Zr^{4+}$ (~0.72 Å) and can alter the local structure, partially stabilizing the higher-symmetry tetragonal and cubic phases at room temperature. Titanium also introduces point defects, particularly oxygen vacancies, depending on processing conditions. Indeed, in zirconia, the tetragonal phase stabilization is not governed only by cation size, but also by oxygen defect chemistry, elastic energy, and local symmetry effects. This compound is interesting because, here again, *ML* appears tied to be linked to lattice distortion or mismatch [8]. The m-$ZrO_2$ ↔ t-$ZrO_2$ transition is first order (martensitic transformation) and the lattice parameters are discontinuous at the interface [37, 38]. In both structures, Zr is seven-fold coordinated and there is one inequivalent Zr site. However, the cation coordination polyhedron in the monoclinic structure becomes increasingly distorted. The monoclinic-to-tetragonal transition with increasing $TiO_2$ content involves a significant jump in several structural parameters and induces a substantial lattice distortion, which gives rise to dislocation-based relaxation mechanisms [39]. In the phase transition region, around 25 mol% $TiO_2$, the monoclinic cell dimensions $a$ and $b$ become almost identical, and $\beta$ drops off significantly, approaching 90° [39]. The maximum *ML* intensity is reported at 0.5 %Ti [40]. The observations of Akiyama et al. [41] were later revisited and reinterpreted in terms of a local piezoelectric field by Chandra [42]. Zr atoms located at the monoclinic–tetragonal interface adopt a strongly distorted [$ZrO_7$] polyhedral structure. A rough estimate, based on the accommodation of the interatomic-distance mismatch over five atomic layers, yields Baur distortion indices $D_s$ in the range 0.12–0.16, significantly larger than in either bulk phase.

e) $ZnS:Mn^{2+}$

Under ambient conditions, cubic ZnS (zinc blende, $F\bar{4}3m$) is the stable phase. The hexagonal wurtzite ($P6_3mc$) phase forms above ~1020 °C and can be retained at room temperature if cooling is fast enough (in a thin-film for example). As $Mn^{2+}$ has a partially filled $3d^5$ configuration, its incorporation into ZnS, where it substitutes for $Zn^{2+}$ at tetrahedral sites, creates localized d-levels within the band gap. This is why Mn-doped ZnS is widely used as phosphor and luminescent material [43-44], with emission typically in the orange range (~585 nm) arising from the ${}^4T_1 \rightarrow {}^6A_1$ transition of $Mn^{2+}$. This substitution induces slight lattice expansion ($r_{Zn}{}^{2+} \approx 0.74$ Å; $r_{Mn}{}^{2+} \approx 0.80$ Å) and local strain. Interest in this compound lies in the fact that, although the cubic form is the most widely studied for luminescence, *ML* is exhibited by the wurtzite structure, which lacks inversion symmetry [42, 45]. More precisely, *ML* in $Mn^{2+}$-doped ZnS is closely associated with a mixed-phase structure of wurtzite (2H) and zinc-blende (3C) ZnS, and is thought to originate from the presence of screw dislocations at the 2H/3C interfaces [46]. As in m-$ZrO_2$/t-$ZrO_2$, the interface between the cubic (ABCABC stacking) and hexagonal (ABAB stacking) phases is a privileged site for stacking faults and can be described by Shockley partial dislocations. These dislocations relieve strain and stabilize the coexistence of two polytypes. Defect-free, single-phase ZnS:Mn nanocrystals without stacking faults do not show strong *ML* [47]. Zn atoms at the cubic–hexagonal ZnS interface exhibit significantly distorted $ZnS_4$ tetrahedra, with $D_s$ in the range 0.04–0.06, compared to values close to zero in the bulk phases.

f) (Ba, Ca)Zn(O, S):($Mn^{2+}$, $Er^{3+}$)

These oxysulfide compounds illustrate the interest of multiple anions for enhancing *ML* performance, possibly through lower symmetry and local disorder at active sites, as suggested by some recent studies [7, 48]. CaZnOS crystallizes in the hexagonal $P6_3mc$ space group, whereas BaZnOS crystallizes in the orthorhombic Cmcm one. When doped with $Mn^{2+}$ or ($Mn^{2+}$, $Er^{3+}$), both phases exhibit *ML*, although the intensity is much weaker for BaZnOS than for CaZnOS [49-52]. On the one hand, Cmcm is centrosymmetric, whereas $P6_3mc$ is hexagonal, non-centrosymmetric, and polar, hence allowing for a piezoelectric effect at rest. On the other hand, $D_s$ is significantly larger for CaZnOS than for BaZnOS (0.148 vs. 0.063). Thus, despite the fact that $Mn^{2+}$ substitution for $Zn^{2+}$ (the ionic radii being close) induces greater lattice distortion in the orthorhombic Cmcm phase, where reduced rotational symmetry and greater structural flexibility permit larger deviations from ideal tetrahedral coordination than the more rigid hexagonal $P6_3mc$ framework, it appears that, in (Ba, Ca)Zn(O, S) compounds, ML is favored by lower overall symmetry, in particular the lack of inversion symmetry, combined with a larger distortion at the Ba or Ca site.

g) BaO-$SiO_2$:($Eu^{2+}$, $Ho^{3+}$)

The two phases listed in Table 1, namely $Ba_4Si_6O_{16}$ ($P2_1/c$) and $Ba_5Si_8O_{21}$ (C2/c), both doped with ($Eu^{2+}$, $Ho^{3+}$) are known luminescent and persistent luminescent phases [53-54]. *ML* was recently reported in both compounds [55, 56], with significantly higher intensity in $Ba_5Si_8O_{21}$:($Eu^{2+}$, $Ho^{3+}$) [55]. It is therefore worthwhile to study these two phases because, although both are centrosymmetric, they nonetheless exhibit *ML*, especially the C2/c phase, which has higher symmetry and is less flexible. Both phases are monoclinic, but the C2/c phase has a larger number of inequivalent $Ba^{2+}$ sites (3 vs. 2) and $O^{2-}$ sites (11 vs. 8). Therefore, since $Eu^{2+}$ substitutes for $Ba^{2+}$, it can occupy multiple sites in the C2/c phase, suggesting a higher degree of disorder than in the $P2_1/c$ phase. Indeed, the distortion at the $Ba_2$ and $Ba_3$ sites in the $Ba_5Si_8O_{21}$ phase is much larger than at any of the barium sites in $Ba_4Si_6O_{16}$ (Table 1).

## III. The driving force: Stress field

### III.1 Macroscale

A given stress field, $\tilde{\sigma}$, consists of a deviatoric part, $\tilde{s}$, and a hydrostatic part, $\tilde{\sigma}_h$: $\tilde{\sigma} = \tilde{s} + \tilde{\sigma}_h$. At the continuum scale, the deviatoric part is responsible for the distortion and induces shape change (for example a square becoming a rhombus), whereas the hydrostatic part drives volume change. The hydrostatic part is defined as: $\tilde{\sigma}_h = (1/3)(\mathrm{Tr}\,\tilde{\sigma})\tilde{1}$, where $\tilde{1}$ represents the unity tensor corresponding to the Kronecker's symbol, and is thus analogous to a pressure, except it can be positive (e.g. in tension) or negative (e.g. in compression). To determine whether a material is more sensitive to shape or volume change, it is interesting to compare the elastic energy stored in each contribution. For a linear elastic material, the total elastic energy stored per unit volume is expressed as: $W_{eV} = (1/2)\ \tilde{\sigma} : \tilde{\varepsilon}$, where $\tilde{\varepsilon}$ is the strain tensor and ":" denotes of doubly contracted product. This elastic energy is the driving force to be considered in the changes of electron energetics responsible for *ML*. The depth of the de-trappable electron states then scales with $\tilde{\sigma}$, so the detrapping rate scales with the mechanical power $\tilde{\sigma} : \dot{\tilde{\varepsilon}}$, where $\dot{\tilde{\varepsilon}}$ is the strain rate [3, 11]. This is why the *ML* intensity scales with mechanical power, accounting for ***KO5***. For constant-rate experiments, the intensity is therefore proportional to the applied stress, and for samples with a constant cross-section area, it is proportional to the load, which is most commonly reported. Following standard derivations [57], the volume densities of energy associated with shape and volume changes are:

$$W_{eV}^{(shear)} = \frac{\tilde{s}:\tilde{s}}{4G} \qquad (3)$$

$$W_{eV}^{(hydro)} = \frac{(Tr\tilde{\sigma})}{18B} \qquad (4)$$

where $G$ and $B$ are the shear and bulk moduli, respectively.

The ratio ($R_{S/V}$) of $W_{eV}^{(shear)}$ to $W_{eV}^{(hydro)}$ compares the elastic energy associated with the shape change (distortion) with that stored through volume change:

$$R_{S/V} = \frac{9B}{2G}\frac{\tilde{s}:\tilde{s}}{(Tr\tilde{\sigma})^2} \qquad (5)$$

The larger $R_{S/V}$, the more shear dominates over pressure and vice-versa, in other words distortion prevails over volume change. $R_{S/V}$ depends both on the loading configuration, and on the material properties. For a pure hydrostatic pressure experiment, for example, $\tilde{s}=\tilde{0}$, so $R_{S/V} = 0$, and the stored elastic energy reduces to $W_{eV} = P^2/(2B)$, where $P$ is the applied pressure ($P=|Tr\tilde{\sigma}|/3$). On the contrary, for a rod or a fiber in torsion (pure shear), $Tr\ \tilde{\sigma} = 0$ and $R_{S/V}$ diverges, while the stored elastic energy reduces to $W_{eV} = \tau^2/(2G)$, where $\tau$ is the shear stress. For a uniaxial compression test, $R_{S/V}$ reduces to:

$$R_{S/V} = 3B/G \qquad (6)$$

Values of $R_{S/V}$ for in uniaxial testing, for the compounds selected in this study that exhibit *EML,* are given in Table 1, together with their elastic properties. Experimentally measured elastic moduli are reported whenever available. When experimental data were unavailable, estimates were obtained from first-principles calculations, performed with the Vienna Ab initio Simulation Package (*VASP*) [58] within the projector-augmented-wave framework [59, 60], using the *PBE*sol generalized-gradient functional [61], which is known to reproduce the lattice parameters and elastic constants of oxide and chalcogenide solids more accurately than standard *PBE*. The polycrystalline bulk and shear moduli ($B$ and $G$) used in equations (3)–(6) were obtained as Hill averages of the Voigt and Reuss bounds derived from $C_{ij}$ [62]. The Voigt average assumes uniform strain across all grains and provides an upper bound on the elastic moduli, whereas the Reuss average assumes uniform stress and provides a lower bound; the Hill average is taken as the arithmetic mean of the two and is generally regarded as the best single estimate for an isotropic polycrystal of randomly oriented grains. Further details on the computational setup are provided in the Supplementary Information.

The reliability of the *PBE*sol functional for the present series can first be assessed at the structural level. Table *S1* (Supporting Information) compares the relaxed conventional cell parameters with the corresponding experimental ones retrieved from the *ICSD*. For all compounds, the deviations on individual lattice constants remain below 1 %, with volume errors typically within 1 %. Slightly larger discrepancies (up to 1.4 % on the cell parameters and ~2 % on the volumes) are observed only for the high-temperature phases of $SrAl_2O_4$ and $BaAl_2O_4$, where 0 K atomic relaxation neglects thermal expansion, accounting for most of the residual error. This level of agreement is in line with the expected accuracy of *PBE*sol on oxides and chalcogenides and validates its use for the determination of the elastic tensors.

For the compounds in this series for which experimental elastic data are available, a direct comparison can be drawn (Table 1). The polycrystalline bulk and shear moduli predicted by *PBE*sol-*DFT* are in very good agreement with the experimental values for cubic ZnS [63] ($B$ = 80 vs 78 GPa, $G$ = 35 vs 33 GPa), tetragonal $Ca_2MgSi_2O_7$ [64] (93 vs 92 GPa for $B$; 35 vs 37 GPa for $G$), and $Ba_4Si_6O_{16}$ [13, 65] (62 vs 58 GPa for $B$; 34 vs 30 GPa for $G$), with relative deviations below 7 %. For the two oxysulfides BaZnOS and CaZnOS, where no experimental measurements have been reported, our *PBE*sol values are consistent with the *DFT* estimates from earlier studies [66, 67], providing an independent cross-check between calculations performed with different functionals and/or codes.

Two cases warrant a closer look. In tetragonal $ZrO_2$, the polycrystalline averages obtained from *DFT* ($B$ = 198 GPa, $G$ = 79 GPa) match the single-crystal Brillouin scattering data of Hayakawa et al. [68] (192 and 82 GPa) within a few percent; however, the individual stiffness constants $C_{44}$ and $C_{66}$ differ markedly. This discrepancy reflects the well-known experimental difficulty of accessing the full elastic tensor of t-$ZrO_2$, which is metastable under stress and tends to transform into the monoclinic phase during single-crystal measurements [69, 70], rather than a failure of the *DFT* calculation. Importantly, since the $R_{S/V}$ descriptor relies on the polycrystalline averages, this discrepancy is inconsequential for the present analysis. In hexagonal ZnS, the *PBE*sol-*DFT* shear modulus (34 GPa) is nearly 50 % larger than the ultrasound value reported by Cline *et al.* [71] (23 GPa), although the bulk modulus is essentially unchanged (78 vs 77 GPa). This discrepancy is consistent with the known sensitivity of wurtzite ZnS samples to stacking faults and polytypism, which softens the apparent shear modulus measured on real polycrystals; the *DFT* value, by contrast, refers to an idealized defect-free crystal and lies within a few percent of the value computed for the cubic polymorph, as expected from the close structural relationship between the two phases. Importantly, the $R_{S/V}$ descriptor is built from the polycrystalline averages $B$ and $G$; for all comparable compounds these averages agree with experiment to within 7 % (with the exception of $G$ for hex-ZnS, discussed above), which supports the use of the *PBE*sol values as reliable proxies for experimental elastic moduli whenever the latter are unavailable.

It turns out that, in a uniaxial test, the elastic energy stored through elastic distortion (change in shape) is generally 5 to 10 times larger than the energy stored through volumetric deformation.

It is worth recalling that i) the $B/G$ ratio can also be used to evaluate the ductility of a material: values above 1.75 indicate a ductile behavior, while lower values indicate brittleness [72], and ii) most materials are more sensitive to shear than to hydrostatic compression, so that it is generally easier to change the shape of a solid than its volume. The trends observed when examining the elastic properties of different compounds are also reflected in Poisson's ratio $\nu$, since $\nu$ increases with the $B/G$ ratio. This is convenient because $\nu$ is known for most compounds and provides a straightforward indication of the connectivity of the atomic network and of the structural dimensionality of the phase [73], a feature that plays a key role on the way external mechanical loading is transfered to the *ML*-active sites. In practice, materials with $R_{S/V} > 7$ can be considered as shear-sensitive at the continuum scale, and typically exhibit easy-shear planes at the atomic scale. Conversely, materials with $R_{S/V} < 6$ display a crosslinked 3*D* character, at least as far as their elastic response is concerned. As expected, $SrAl_2O_4$ falls in the latter category. More surprisingly, this is also true for the barium silicates investigated here, despite the well-known presence of easy-shear barium-rich layers in their structures. The topology of these layers, in particular their tortuosity and finite extent, likely explains why the surrounding silicate framework still imparts a 3*D* character to the macroscopic elastic behavior. Then, although $R_{S/V}$ values above 6 could be anticipated for the melilite-type crystals (6.2 and 8.0 for A = Ba and Ca, respectively in $A_2MgSi_2O_7$), given the easy shear of their basal planes, the large values (>7) observed for cubic ZnS and tetragonal zirconia are more counter-intuitive, since both crystals possess a pronounced 3*D* character. ZnS and $ZrO_2$ also differ from the other compounds of Table 1 in a fundamental way: they are binary phases featuring a single type of chemical bonding, whereas all the other materials are multinary and combine several types of cation–anion bonds. The competition between these bonds, which are all predominantly iono-covalent in nature, is therefore expected to influence the elastic response. Using Pauling electronegativity differences as a proxy for ionicity, the cation–anion bonds present in our series can be ranked as follows: Ba–O (2.6) > Sr–O (2.5) > Ca–O, Mg–O, Zr–O (2.2) > Al–O (1.9) > Si–O (1.6) > Zn–S (0.8); the Allred–Rochow electronegativity scale yields a very similar ordering. ZnS thus stands out as the most covalent compound of the series, while the alkaline-earth aluminates and silicates exhibit a mixed iono-covalent character with simultaneous contributions from highly ionic *A*–O bonds (*A* = Ba, Sr, Ca) and more covalent Si–O or Al–O bonds. This is mechanically relevant because the more covalent a bond, the more directional it is, and the more strongly it resists angular distortion, and hence shear; conversely, the highly ionic *A*–O bonds (and especially Ba–O) are essentially non-directional and allow easy gliding of the surrounding atomic layers. The high $R_{S/V}$ values found for the Ba-rich silicates and aluminates can therefore be rationalized, at least in part, by the prevalence of weakly directional Ba–O contacts in their structures, whereas the strong directionality of Si–O bonds maintains the integrity of the surrounding silicate framework.

It also should be noted that the values of $\nu$ and $R_{S/V}$ listed in Table 1 do not, by themselves, allow firm conclusions to be drawn: they correspond to macroscopic averages for an isotropic polycrystal, whereas the elastic response along specific directions can deviate substantially from these averages, as discussed in § III.2. In structures exhibiting 2*D* regions (such as the Ba and Ca melilites and the barium silicates), shear can be facilitated along specific crystallographic planes and directions, while the interlayer distance varies little, with only a minor incidence on the crystal-field strength at the active sites. Shear may also occur in regions of the structure that do not host the active centers. In such cases, the hydrostatic component of the stress is expected to play an important role. In contrast, in 3*D*-interconnected structures (such as $SrAl_2O_4$), shear is distributed throughout the crystal and necessarily alters the crystal-field strength at the active sites, irrespective of the shear direction, and during both loading and unloading stages. A large $R_{S/V}$ ratio combined with a 3*D* directional-bonding network is therefore expected to be most favorable for triggering *ML*, whereas in 2*D* structures the hydrostatic part of the stress may dominate the response, a feature that contributes to explaining the behavior of ***KO6***.

Now, given that shear plays an equivalent role in tension and in compression, whereas the hydrostatic component of the stress is negative in compression and positive in tension, the shear contribution emerges as a natural candidate to explain ***KO4***, namely, the occurrence of light emission during both loading and unloading. Hydrostatic pressure is unlikely to play a major role here, since the crystal-field strength ($\Delta_0$) deceases rapidly upon volume expansion and increases upon volume contraction ($\Delta_0 \propto 1/r_0^5$, where $r_0$ is the metal-ligand distance [74]), unless pressure also induces significant shear at the crystal-structure scale. This may explain why a *ML* peak is observed both during loading and unloading for $SrAl_2O_4{:}Eu^{2+}$ in uniaxial compression, that is, in the presence of a non-zero deviatoric component, whereas the unloading peak vanishes under purely hydrostatic conditions. Nevertheless, the *ML* performance can hardly be correlated to $R_{S/V}$ alone, because the way macroscopic distortion (or, equivalently, the deviatoric elastic energy stored in the polycrystalline material) is transferred to the active sites must to be taken into account. A key requirement is that shear actually reaches the sites relevant to the *ML* process, and this is where the crystal structure and its associated elastic anisotropy come into play.

### III.2 Structural scale (single crystal)

At the crystal scale, it is instructive to examine the elastic peculiarities of *ML* phases and the manner in which they deform under load. The elastic response of a single crystal is anisotropic, and the linear elastic constitutive relations write:

$$\sigma_I = \sum_{J=1}^{6} C_{IJ}\varepsilon_J \qquad (7)$$

and conversely,

$$\varepsilon_I = \sum_{J=1}^{6} S_{IJ}\sigma_J \qquad (8)$$

where $\sigma_I$ and $\varepsilon_I$ are the usual stress and strain components in Voigt (contracted) notation, expressed as six-component vectors for stress and strain.

The elasticity constants $C_{IJ}$ and $S_{IJ}$ are the components of the anisotropic stiffness and compliances tensors, respectively [62]. Recall that for *I* (or *J*) = 4, 5 or 6, $\varepsilon_I$ correspond to an angular distortion, and is therefore equal to twice the conventional tensorial strain component, for instance, $\varepsilon_5 = \gamma_{13} = 2\varepsilon_{13}$. Both $C_{IJ}$ and $S_{IJ}$ tensors are symmetric. Their number of independent components ranges from 3 in cubic system to 21 in the triclinic case. The hexagonal, tetragonal, orthorhombic, and monoclinic systems, which are of particular interest in the present work, are characterized by 5, 6, 9 and 13 independent constants, respectively. It is worth noting that $S$ is the relevant tensor for investigating the orientation dependence of the elastic moduli: for instance, the inverse of Young's modulus along an arbitrary direction $\vec{n}$ is given by $1/E(\vec{n}) = S_{ijkl}n_i n_j n_k n_l$ . As a consequence of elastic anisotropy, distortion preferentially develops along specific planes and directions of the crystal. However, it is noteworthy that hydrostatic pressure itself can induce distortion at the polyhedral scale: pressure induced structural transitions involving inter-polyhedral bond-angle changes have been documented for silicate glasses [75-78] or perovskite crystals [79], and provide direct evidence of such effects (Fig. 3a). Furthermore, crystals of lower symmetry, and in particular the monoclinic phases studied here, exhibit pronounced couplings between normal and shear stresses, manifested by non-zero values of the off-diagonal constants $C_{15}$, $C_{25}$ and $C_{35}$ (see Supplementary Materials).

In 3*D*-interconnected structures where the active sites (luminescence centers and traps) are embedded in a network of relatively stiff polyhedral units (e.g. tetrahedra, octahedra), a global deformation of the crystal is required to induce a local deformation in the vicinity of the active sites. This situation is encountered in both the monoclinic and hexagonal $SrAl_2O_4$ phases, where the $Sr^{2+}$ ions occupy relatively large sites surrounded by layers of six-membered rings of corner-sharing [$AlO_4$] tetrahedra (Fig. 3b). In such structures, shear is most easily activated along the basal *(ab)* planes parallel to the rings, or along (*bc)* planes in the *b* direction, but the three shear moduli of the crystal differ by less than 30 %. Figure 3b illustrates a shear deformation along the (*bc*) planes, inducing an angular change of 20° (associated with $\varepsilon_6$) in the *ab* plane and corresponds to a shear stress $\sigma_6 = C_{66}\,\varepsilon_6$ of about 17 GPa. Under this loading, the regular hexagonal rings are deformed into distorted ones, and the planar value of the distortion index ($D_s$) increases from zero to about 0.1. A shear stress of ~10 GPa would induce a distortion ($D_s$ = 0.059) comparable in amplitude to the intrinsic structural distortion of the $SrAl_2O_4$ phases reported in Table 1. Likewise, the monoclinic-to-orthorhombic transformation, closing the angle $\beta$ from 93.42° to 90°, requires a shear strain $\varepsilon_5$ = tan(3.42°) $\approx$ 0.060, which corresponds to a shear stress $\sigma_5 = C_{55}\varepsilon_5 \approx$ 3.3 GPa, with $C_{55} \approx$ 55 GPa (Table 1). A similar mechanism may operate in $Ba_5Si_8O_{21}$ and, to a lesser extent, in $Ba_4Si_6O_{16}$, both of which exhibit $R_{S/V} < 6$. In the case of BaZnOS and CaZnOS (the latter being *EML*-active*)*, the substantially higher distortion index of the calcium phase appears to prevail over the small difference in $R_{S/V}$ between the two compounds. Finally, it should be noted that in a polycrystalline ceramic, an angular distortion induced by a non-zero $\sigma_5$ component is statistically as likely to close the $\beta$ angle as to open it (Fig. 4b), regardless of the sign of $\sigma_5$. This symmetry provides an additional argument for the appearance of an *ML* signal during both the loading and unloading stages, since the sign of the shear stress reverses between the two.

In structures with a pronounced 2*D* character, sliding is facilitated along specific crystallographic planes and occurs while leaving rest of the structure virtually undistorted away from these planes. This is the case for the melilite crystals ($A_2MgSi_2O_7$, where $A$ = Ba, Sr, Ca), which consist of layers of relatively rigid polyhedral interleaved with large, irregular cavities hosting the $A^{2+}$ cation; here too, shear is most easily activated along the basal planes (Fig. 3c). In $Ca_2MgSi_2O_7$ melilite, the Voigt-Reuss-Hill average of Poisson's ratio is $\nu$ = 0.32 [64], but the directional values are markedly anisotropic: $\nu_{13} = -S_{13}/S_{11} \approx 0.34$ ($= -\varepsilon_c/\varepsilon_a$ for a tension along $a$) is much smaller than $\nu_{31} = -S_{13}/S_{33} \approx 0.40$ ($= -\varepsilon_a/\varepsilon_c$ for a tension along $c$), reflecting the equivalence of the $a$ and $b$ axis and the noticeably different response along c (Fig. 3c). As a consequence, a tensile loading in the *ab* plane induces only a "limited" contraction of the interlayer spacing along $c$, whereas a tension along the $c$ produces a comparatively large transverse contraction along $a$ and $b$. In this compound, shear is most likely to develop along *ab* planes, with little change in the interplane spacing and, consequently, only modest variations of the crystal-field strength at the Ca sites. The O-Ca-O layers are indeed substantially more compliant than the surrounding [$SiO_4$] and [$MgO_6$] polyhedral layers. In order to have an idea, the bonding-energy ratio between four-coordinate Mg-O and seven-coordinate Ca-O bonds is about 1.64 and the $S_{55}$ component of the compliance of the P-$42_1$m tetragonal phase is roughly twice that of $S_{44}$. In this phase, the $Eu^{2+}$ activator predominantly substitutes for $Ca^{2+}$. Since interlayer shear is expected to perturb bond lengths only weakly compared to hydrostatic pressure, its impact on *ML* intensity is anticipated to be smaller than that of pressure, a behavior that should also apply to the barium silicates of Table 1. For $Ba_4Si_6O_{16}$, whose structure consists of layers of [$SiO_4$] tetrahedra (stiff units) separated by $Ba^{2+}$ ions, it has indeed been shown that shear plays a negligible role on *ML,* while hydrostatic pressure is the dominant mechanism. This is corroborated by the absence of an unloading *ML*-peak in this compound, and suggests that the active sites (charge carrier traps; likely oxygen vacancies) are essentially insensitive to shear. By comparison, $Ba_5Si_8O_{21}$ (space group C2/c) is somewhat less compact and less stiff than $Ba_4Si_6O_{16}$ and is therefore more prone to volume changes, as also reflected in its lower value of $R_{S/V}$. The superior *EML* performance of $Ba_5Si_8O_{21}$ thus likely originates from the existence of a relatively distorted barium site, with $D_s = 0.054$ for the $Ba_3$ site, to be compared with values close to 0.03 in $Ba_4Si_6O_{16}$.

A simple calculation based on the elastic constants of Table 1 shows that the distortions induced by shear stresses in the GPa range are comparable in magnitude to the intrinsic structural distortions of these compounds, whereas stresses in the 100 MPa range, typical of standard mechanical testing, with the noticeable exception of sharp-contact experiments such as scratch or indentation, induce distortions that are 10 to 30 times smaller, as further discussed in §. IV. For instance, in $Ca_2MgSi_2O_7$ melilite, a shear stress $\sigma_4$ of 100 MPa would induce an angular distortion $\varepsilon_4 \approx 3.3\ 10^{-3}$, corresponding to a Baur distortion index of about 1.7 $10^{-3}$, roughly 35 times smaller than the intrinsic structural distortion ($D_s$=0.058).

**III.3 Atomic scale: Where the electrons come into play**

Several models have been proposed to account for the *EML* process at the atomic scale. These models invoke changes in the energetics of the electrons and consider, alternatively: (i) the transfer of an electron from the luminescence center to the conduction band, followed by trapping and eventually recombination at the luminescent center; (ii) the separation of electron–hole pair under a piezoelectric field, with the electron promoted to the conduction band and the hole to the valence band, followed by recombination; or (iii) the appearance of polarization charges upon mechanical loading, which de-trap previously trapped carriers via tunneling mechanism, with subsequent recombination at the luminescent centers. These models are reviewed in [3, 6]. The question that remains open is how the applied stress field is converted into a driving force at the active site, and what stress intensity, orientation and components (shear or hydrostatic) are required to activate each of the above mechanisms. For instance, in the trapping-detrapping picture, the driving force must be sufficient either to release the trapped electrons or to perturb the conduction- or valence-band edges meaningfully. In other words, the stress-induced reduction of the activation energy in the classical persistent-luminescence relation [80] must be large enough for detrapping to occur on a typical experimental timescale. The local change in the crystal-field strength, in turn, depends on the orientation of the active electronic orbitals relative to the local strain field. These two fundamental aspects of the crystal-to-atom-scale transition lie beyond the scope of the present study.

At the crystallographic level, and in particular as regards the local coordination environment of the active sites, the distortion index, $D_s$, appears at first sight as an attractive descriptor of the local geometry. In the recent work by Uchiyama [18] and Lin [7], $D_s$ indeed emerged as a relevant proxy for *EML* performance, and our comparative analysis (Table 1) also suggests that a large $D_s$ value is favorable for *EML*. However, $D_s$ is most often computed from the bond lengths within the polyhedron centered on the host cation (Sr in $SrAl_2O_4$, for instance), whereas the luminescent activator most often substitutes for that host cation; furthermore, the index does not account for the orientation of the mechanical loading. In the specific context of *ML* studies, we therefore suggest that $D_s$ be (i) weighted by the number of presumably most active sites, (ii) further normalized to a unit of volume to allow direct comparison across compounds, and possibly (iii) restricted to bond directions relevant to the crystal symmetry and the loading axis. It is also interesting to consider the contribution to the total distortion stemming from the mechanical loading. For instance, in the simple cases illustrated in Figs 3b and 2c, a shear-stress in the GPa range produces $D_s$ values comparable to those of the intrinsic structural distortion at rest. The dependence of $D_s$ on the applied stress is examined for a few representative geometries in the following section.

## IV. Consequences: Kinematics of the crystal deformation

The way a crystal deforms under a given mechanical loading, and, in particular, the way the local environment of the luminescent centers is affected, is of paramount importance for understanding the underlying *EML* mechanisms. Because minor changes of the metal-ligand distances have a strong incidence on the crystal-field strength ($1/r_0^5$ scaling discussed earlier), it is essential to assess how interatomic distances or local volumes evolve at active sites.

Several representative cases are illustrated in Fig. 4. In order to highlight the essential geometrical features of the distortion mechanisms, these cases are represented in a simplified two-dimensional fashion, in which the polyhedra are not drawn explicitly: each node represents the center of a rigid structural unit (tetrahedron or octahedron), and the connecting segments stand for the angular articulations between neighboring polyhedra (as in Figs. 3b and 3c). Fig 4a depicts four corner-sharing octahedra, a motif typical of $ABC_3$ perovskites ($BaTiO_3$, $LaGaO_3$, $CsPbBr_3$, etc.), under shear, hydrostatic pressure, and combined shear-and-pressure loading. It shows how shear distorts the central cavity (as in Fig. 3a), while hydrostatic pressure changes its volume but preserves its shape, except when applied to an already distorted cavity, thus exacerbating the distortion. The latter situation has been observed experimentally for $CsPbBr_3$ perovskite (Fig. 3a). It is also worth noting that, although shear globally reduces the volume of the cavity, it elongates one of its dimensions while shortening the other. This local symmetry breaking, which accounts for ***KO10***, reflects in the increase of $D_s$ with distortion or upon shearing. The relation between cavity distortion and changes in the crystal-field strength is, however, not straightforward. The case of a monoclinic phase under shear is shown in Fig 4b, for two crystals with opposite orientations relative to the loading direction. While the first crystal experiences a decrease of the angle $\beta$, the second experiences the opposite, and vice-versa when the shear direction is reversed. This geometrical observation explains, on one hand, why some compounds are far more sensitive to shear than to hydrostatic pressure (***KO6***) and show *EML* either on loading and unloading (***KO4***). On the other hand, it also provides a natural rational for the differences in *EML* response observed when the sample is irradiated under load (***KO7***). Indeed, assuming that the crystal-field strength scales as an inverse power of the metal-ligand distance, and therefore also with volume of the cavities hosting the active sites in Fig. 4b, and assuming further that a larger volume corresponds to a shallower trap, it follows that different traps are populated depending on whether the sample is at rest or under load. This addresses ***KO7***, and the corresponding response during successive loading cycles will therefore differ, as required by ***KO4***. The same schematic also suggests that the trapping efficiency is maximal at a particular value of the distortion (corresponding to a maximal cavity volume), which may be at the origin of ***KO8***. Such a behavior is expected in 3*D*-interconnected structures, where the whole crystal is necessarily strained under load, as in the Ba- or Sr-aluminates discussed above. In contrast, in 2*D* (layered) structures, such as the Ba- and *RE*-silicates, shear can be confined to easily shearable regions, leaving the rest of the crystal almost undeformed. In this configuration, illustrated in Fig. 4c, *EML* can only be observed if the active sites are themselves located within the distorted regions, and are therefore sensitive to the applied stress field. This situation occurs, for instance, at grain boundaries between crystallites and in melilite (Fig. 3c) or phyllosilicates structures. A more systematic investigation of the dependence of $D_s$ on the applied shear was carried out for a set of simple polygonal motifs (triangle, square, hexagon and octagon) by applying the Baur expression of Eq. (2) to their distorted forms, as a function of the distortion angle (Fig. 5). The results show that $D_s$ is essentially proportional to the distortion angle for distortions up to $\theta = 20°$ (Fig. 5b). Since the elastic constants of the compounds studied here are of the order of a few tens of GPa, an applied stress in a few tens of MPa range corresponds to a strain of the order of $10^{-3}$, hence to a distortion angle below 1°, which yields $D_s$ values in the 0.002-0.01 range, that is, slightly below the typical structural-distortion values reported for *EML* compounds, but of the same order of magnitude. Conversely, in high-pressure experiments (above 1 GPa) or under indentation- and scratch-test conditions, where the contact stresses can reach several GPa (the hardness scale), the mechanically induced distortion may dominate over the intrinsic structural one.

These preliminary considerations illustrate the complexity of the problem. Hydrostatic pressure and shear produce markedly different consequences depending on the crystal structure, not to mention indentation- and scratch-test geometries, which generate highly complex stress states reaching the GPa range [81, 82]. In some cases, the applied stress field may simultaneously decrease the local crystal-field strength and increase the trap depth. Furthermore, at the discrete scale of the atomic bonding network, shear may induce volume changes while hydrostatic pressure may induce shape changes, in apparent contradiction with the classical continuum-scale picture.

## V. Conclusion and perspectives

Science progresses by asking questions. We hope this article will encourage further investigations focused on the fundamental understanding of *EML*, eventually leading to a coherent physical picture and to quantitative model(s) based on a clear description of how an applied stress field affects the local crystal-field strength at the active luminescent and charge carrier trapping sites. In our view, future studies should pay particular attention to the correlations between (i) the characteristics of the stress field (orientation, hydrostatic versus shear contributions), (ii) the resulting local distortion, and (iii) the local crystal-field strength. A better understanding of these correlations cannot be achieved without (a) systematic comparisons of the behavior under shear (or combined shear and pressure, as in uniaxial tests) and under purely hydrostatic loading, with *in-situ* structural monitoring whenever possible, and (b) the determination of the anisotropic elastic tensor of the phases, together with an analysis of how the external stress field is transferred to the active sites at the atomic scale. The structural distortion index, whose static nature does not fully reflect the actual situation at the onset of mechanoluminescence, should ideally be supplemented by an estimate of the dynamic distortion induced by the loading. A simple analysis based on a few motifs representative of the mechanically induced distortion associated with *EML* experiments is, in many cases, comparable to that of the intrinsic structural distortion at rest.

**Acknowledgements**: One of the authors wishes to thank the University of Kyoto for its support of this research, particularly through a visiting professorship. We extend our sincere thanks to Alexis Duval and Xu Jian for their stimulating discussions and careful review of the manuscript.

## Declaration of competing interest

The authors declare that they have no known competing financial interests or personal relationships that could have appeared to influence the work reported in this paper.

## Appendix A. Supplementary data

**Table 1** – *PBE*sol-*DFT* elastic properties of the studied compounds, compared with available experimental data. For each compound, the upper line gives the experimental reference (when available) and the lower line the present *PBE*sol-*DFT* calculation. The bulk modulus $B$, the shear modulus $G$ and the Poisson ratio $\nu$ are Hill averages of the Voigt and Reuss bounds. The shear stiffness coefficients $C_{44}$, $C_{55}$ and $C_{66}$ are given in Voigt notation (GPa); the remaining coefficients can be found in the referenced papers and in Supplementary Materials. The shear-to-volume energy ratio $R_{S/V}$ was computed from Eq. (6). The distortion index $D_s$ of the relevant cation polyhedron (*CN*: coordination number) was calculated from the expression of Baur [19] (Eq. (2)) using the structural information from the corresponding *CIF* files, by means of a Python script developed for this purpose. Methods abbreviations: *DFT* - Density Functional Theory; *MR* - Mechanical Resonance technique; *US* - Ultrasound technique; *XR* - X-Ray diffraction technique. *nd*: not determined. *Estimated from an almost fully crystallized glass-ceramic.

| Compound | Crystal system | G (GPa) | B (GPa) | ν | $C_{44}$, $C_{55}$, $C_{66}$ (GPa) | Method | $R_{S/V}$ = 3B/G | $D_s$ (CN) | Ref. | Observations |
|---|---|---|---|---|---|---|---|---|---|---|
| $SrAl_2O_4$ | Hexagonal ($P6_3$) | 44 | 84 | 0.28 | $C_{44}=C_{55}=45$, $C_{66}=40$ | *DFT* | 5.7 | Sr(9): 0.056 | This work | *3D* interconnected framework of $[AlO_4]$ |
| | Monoclinic ($P2_1$) | 45 | 78 | 0.260 | $C_{44}=39$, $C_{55}=56$, $C_{66}=50$ | *DFT* | 5.3 | $Sr_1$(7): 0.065<br>$Sr_2$(8): 0.077 | This work | |
| $BaAl_2O_4$ | Hexa. high T ($P6_322$) | 28 | 91 | 0.36 | $C_{44}=13$, $C_{55}=13$, $C_{66}=36$ | *DFT* | 9.7 | Ba(9): 0.017 | This work | Easy shear in basal planes |
| | Hexa. low T ($P6_3$) | 42 | 90 | 0.30 | $C_{44}=43$, $C_{55}=43$, $C_{66}=38$ | *DFT* | 6.5 | $Ba_1$(9):0.058<br>$Ba_2$(9): 0.027 | This work | Ba channels in a framework of corner sharing $[AlO_4]$ |
| $Ba_2MgSi_2O_7$ | Tetragonal ($P\text{-}42_1m$) | 45 | 92 | 0.29 | $C_{44}=C_{55}=47$; $C_{66}=51$ | *DFT* | 6.2 | Ba(8): 0.018 | This work | 2D struct.; easy shear along ab planes ($C_{44}$&$C_{55}<C_{66}$) |
| $Ca_2MgSi_2O_7$ | Tetragonal ($P\text{-}42_1m$) | 37<br>35 | 92<br>93 | 0.32<br>0.33 | $C_{44}=C_{55}=30$, $C_{66}=59$<br>$C_{44}=C_{55}=23$, $C_{66}=60$ | *MR*<br>*DFT* | 7.5<br>8.0 | Ca(8): 0.058 | [64]<br>This work | 2D struct. $C_{44} \neq C_{66}$ (strong anisotropy) |
| $ZrO_2$ | Tetragonal ($P4_2/nmc$) | 82<br>79 | 192<br>198 | 0.32<br>0.32 | $C_{44}=C_{55}=65$, $C_{66}=70$<br>$C_{44}=C_{55}=36$, $C_{66}=170$ | *XR*<br>*DFT* | 7.0<br>7.5 | Zr(8): 0.071 | [68]<br>This work | Distortion at the m-$ZrO_2$/t-$ZrO_2$ interfaces. |
| | Monoclinic ($P2_1/c$) | 89 | 174 | 0.28 | $C_{44}=87$, $C_{55}=86$, $C_{66}=131$ | *DFT* | 5.8 | Zr(7): 0.028 | This work | |
| ZnS | Cubic ($F\text{-}43m$) | 33<br>35 | 78<br>80 | 0.32<br>0.31 | $C_{44}=C_{55}=C_{66}=46$<br>$C_{44}=C_{55}=C_{66}=49$ | *US*<br>*DFT* | 7.1<br>6.8 | Zn(4): 0 | [63]<br>This work | Screw dislocations in the 2H (wurtzite) / 3C (zinc-blende) interfaces. |
| | Hexagonal ($P6_3mc$) | 23<br>34 | 77<br>78 | 0.37<br>0.31 | $C_{44}=C_{55}=29$, $C_{66}=33$<br>$C_{44}=C_{55}=29$, $C_{66}=34$ | *US*<br>*DFT* | 10.2<br>6.9 | Zn(4):≈ 0 | [71]<br>This work | |
| BaZnOS | Orthorhombic ($Cmcm$) | 37<br>39 | 61<br>68 | 0.25<br>0.26 | $C_{44}=47$, $C_{55}=29$, $C_{66}=30$<br>$C_{44}=27$, $C_{55}=31$, $C_{66}=56$ | *DFT*<br>*DFT* | 5.0<br>5.3 | Ba(8): 0.063 | [66]<br>This work | *3D* interconnected network, despite some elastic anisotropy |
| CaZnOS | Hexagonal ($P6_3mc$) | 35<br>36 | 75<br>78 | 0.30<br>0.30 | $C_{44}$=$C_{55}$=30, $C_{66}$=46<br>$C_{44}$=$C_{55}$=34, $C_{66}$=39 | DFT<br>DFT | 6.4<br>6.5 | Ca(6): 0.148 | [51]<br>This work | 3D structure with anisotropic shear ($C_{66}$ > $C_{44}$=$C_{55}$) |
| $Ba_4Si_6O_{16}$ | Monoclinic ($P2_1/c$) | 30*<br>34 | 58*<br>62 | 0.28*<br>0.27 | *nd*<br>$C_{44}$≈$C_{55}$≈$C_{66}$≈36 | US<br>DFT | 5.8*<br>5.5 | $Ba_1$(8): 0.031<br>$Ba_2$(8): 0.027 | [13, 65], This work | 1*D*-2*D* struct. with limited elastic anisotropy |
| $Ba_5Si_8O_{21}$ | Monoclinic ($C2/c$) | 32 | 56 | 0.26 | $C_{44}$≈$C_{55}$≈$C_{66}$≈34 | DFT | 5.2 | $Ba_1$(8): 0.021<br>$Ba_2$(8): 0.028<br>$Ba_3$(8): 0.054 | This work | 1*D*-2*D* struct. with limited elastic anisotropy; less packed and lower $R_{S/V}$ than $Ba_4Si_6O_{16}$ |

Figures

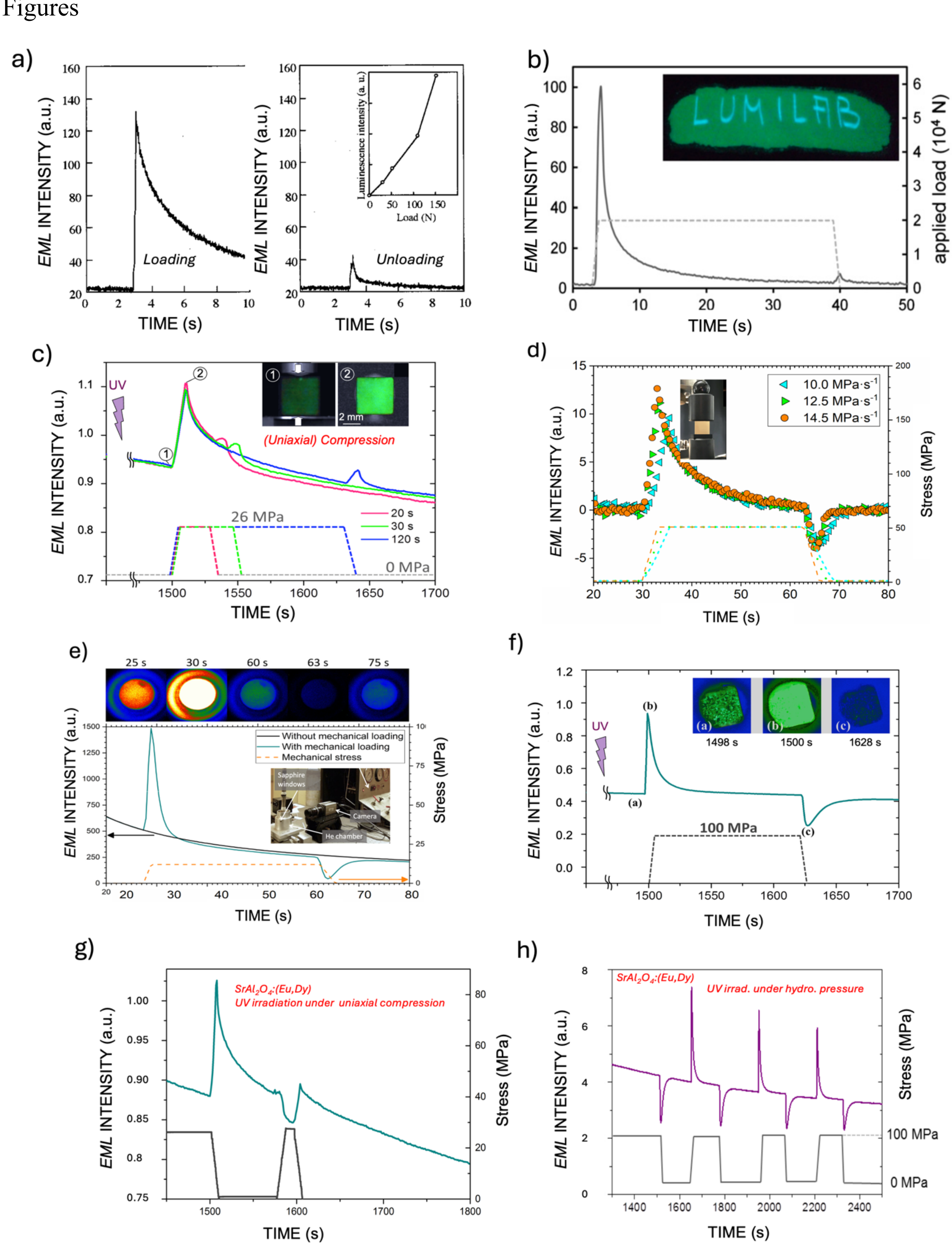

Fig. 1 – *EML* observations in different materials during loading and unloading, under various loading conditions: a) *EML* peaks observed during loading and unloading of $ZrO_2$:$Ti^{4+}$ crystals (powder/epoxy composite) [8]; b) *EML* peaks observed during loading and unloading of a $BaSi_2O_2N_2$:$Eu^{2+}$ powder/epoxy composite under 64 MPa uniaxial compression [10]; c) *EML* peaks observed during loading and unloadingof a $SrAl_2O_4$:($Eu^{2+}$, $Dy^{3+}$) glass/particulate composite under uniaxial compression, for different loading durations [11]; d) decrease of the *EML* intensity upon unloading in a $Ba_2Si_3O_8$:($Eu^{2+}$,$Ho^{3+}$) glass-ceramic containing 4 at.% N [Duval, 2023]; e) Intense *EML* signal under hydrostatic pressure in a $Ba_2Si_3O_8$:($Eu^{2+}$, $Ho^{3+}$) glass-ceramic [Duval, 2023]; f) disappearance of the unloading *EML* peak upon decompression of $SrAl_2O_4$:($Eu^{2+}$, $Dy^{3+}$) crystals in a He-gas chamber [12]; g) loading/unloading cycle in uniaxial compression following *UV* irradiation under load in $SrAl_2O_4$:($Eu^{2+}$, $Dy^{3+}$); and h) loading/unloading cycles under hydrostatic pressure following *UV* irradiation under pressure in $SrAl_2O_4$:($Eu^{2+}$, $Dy^{3+}$) [15].

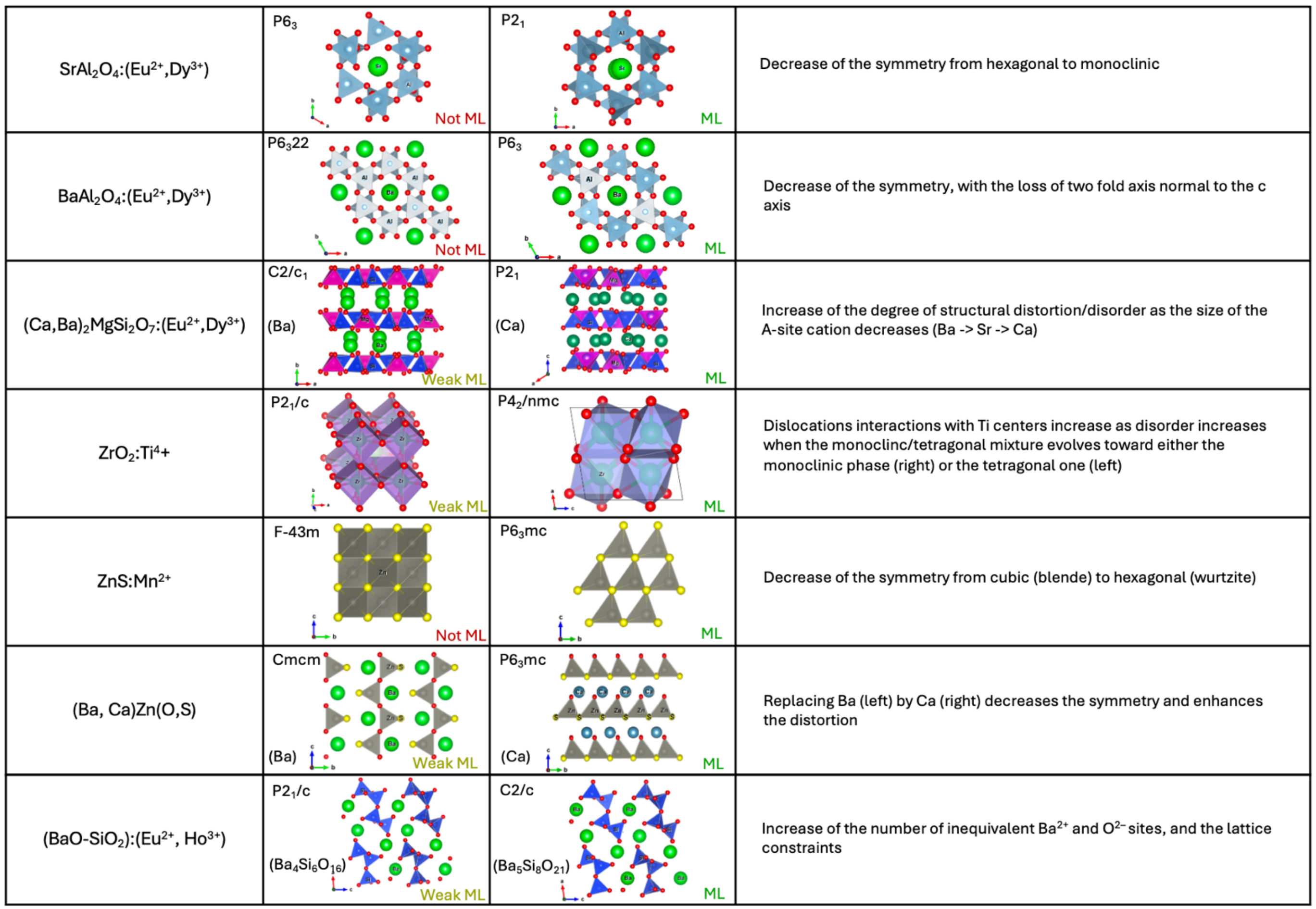

Fig. 2 – Different chemical systems in which *EML*-active compounds have been reported. The first column lists the system. The second and third columns compare selected compounds within each system, most often polymorphs, that exhibit weak or undetectable *EML* signals (second column) and, conversely, those for which a clear and relatively intense *EML* signal has been established (third column). Cases of cationic substitutions (for example Ca for Ba) or minor changes in stoichiometry, as in the BaO-$SiO_2$ system (last row), are also considered. The fourth column summarizes the structural or chemical features that can be invoked to rationalize the *EML* enhancement observed between the compounds in the second and third columns. References to the corresponding studies are given in the main text (§ II.2).

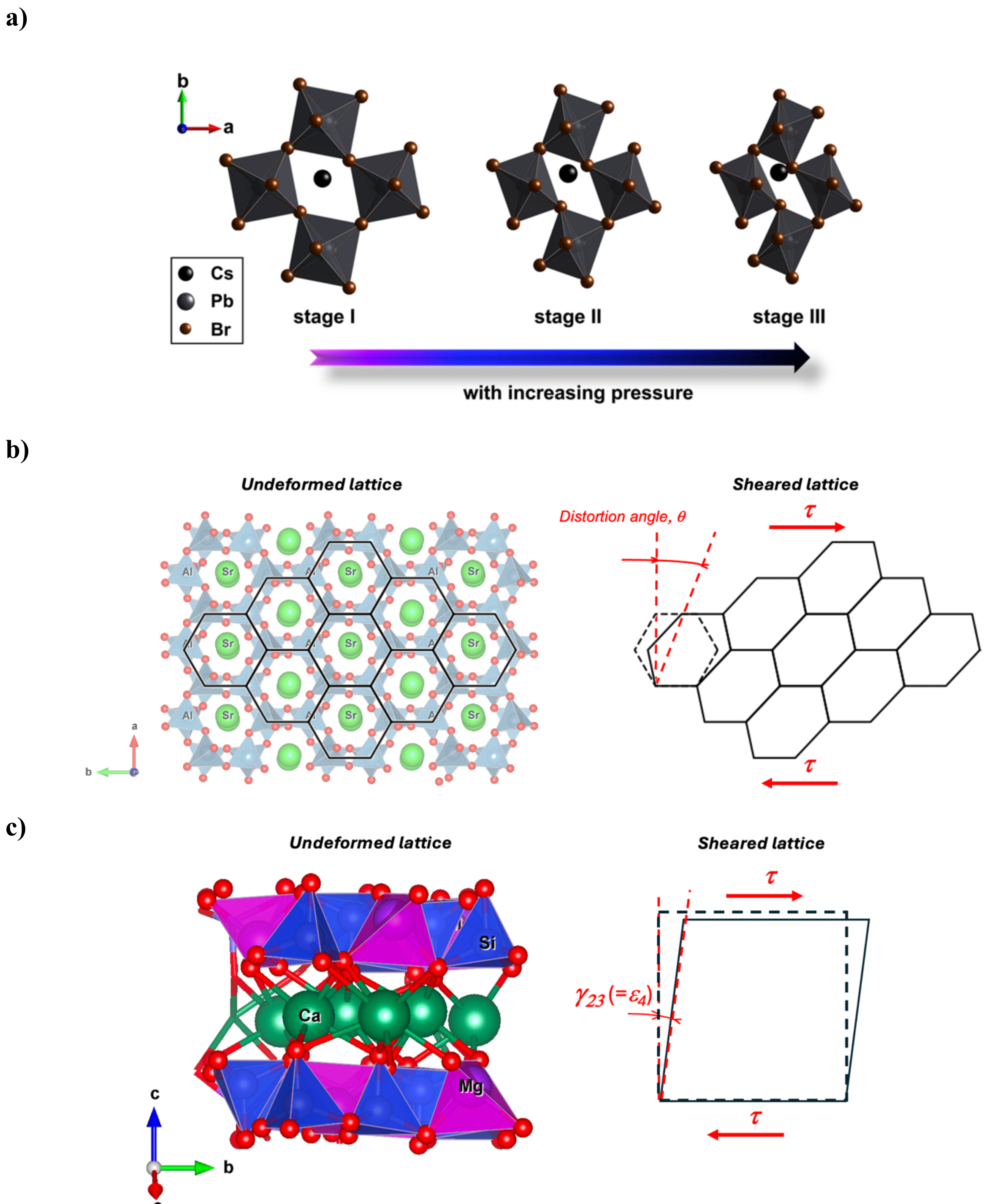

Fig. 3 – a) Distortion of the Pb-containing cavity in $CsPbBr_3$ perovskite under high pressure (reprinted from [79]); b) Shear-induced distortion of the monoclinic lattice of $SrAl_2O_4$ along the *(bc)* planes in the *b* direction, resulting in a $\gamma_{12}$ distortion (between *a* and *b* axes), where for sake of clarity a distortion angle of 20° was chosen, corresponding to a shear stress of about 17 GPa ($=C_{66}\varepsilon_6$). In this case, $D_s \approx 0.1$ (planar estimate); and c) Shear of a $Ca_2MgSi_2O_7$ crystal along the basal planes in the *b* direction, inducing a *bc* angle distortion ($\gamma_{23}$). The angular distortion is of 10° and corresponds to a distortion of 0.176, induced by a shear stress of about 5.2 GPa ($C_{44}\varepsilon_4$). In this latter case, $D_s \approx 0.008$.

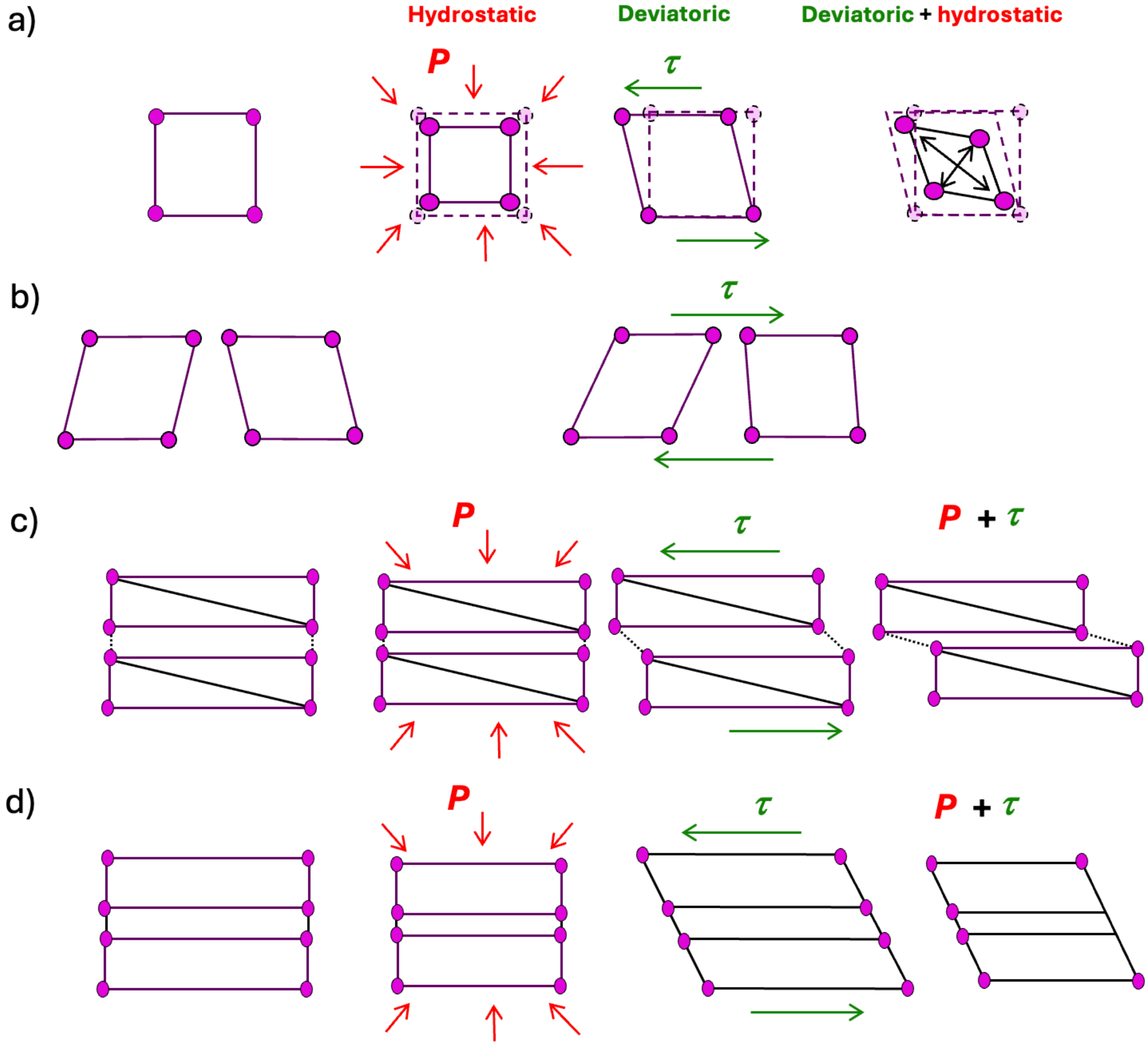

Fig. 4 – Schematic two-dimensional representations illustrating the kinematics of crystal deformation in several typical situations: a) Although hydrostatic pressure is expected to change the volume of the cavity while leaving its shape invariant, when superimposed on shear or applied to a structure that is already distorted, pressure exacerbates the distortion; b) The effect of shear on a monoclinic polycrystal depends on whether the angle $\beta$ increases or decreases under load: a volume contraction is induced in the first case, whereas an expansion is observed in the second case; both situations occur with equal probability in an isotropic polycrystalline aggregate; c) The case of a $2D$ (layered) structure with a relatively compliant interphase, showing that shear is confined to the layered region and leaves the surrounding environment essentially undeformed, in contrast with d) where a stiff interface comparable to the bulk crystal results in a more homogeneous distortion of the whole crystal.

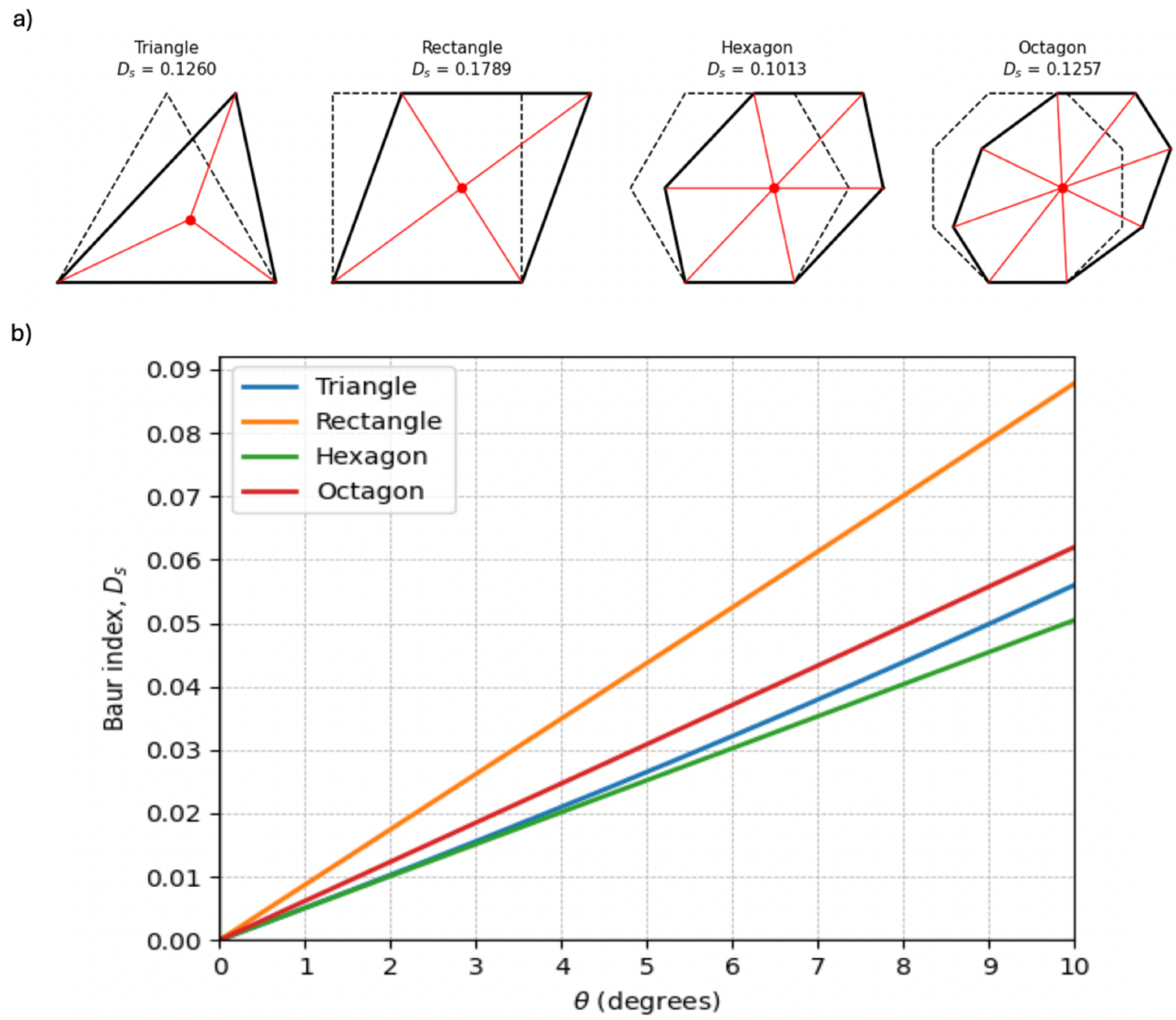

Fig. 5 – Evolution of the Baur distortion index, $D_s$, as a function of the angular distortion ($\theta$). a) schematic illustrations of four different polygonal motifs subjected to a horizontal shear $\theta$ = 20 °; b) Variation of $D_s$ with $\theta$ for $\theta$ up to 10°, for each of the four motifs.

# Supplementary materials: Mechanoluminescence in crystalline inorganic materials: local disorder and the elastic distortion hypothesis

T. Rouxel[1,2], X. Rocquefelte[3] and S. Tanabe[1]

[1]Graduate School of Human and Environmental Studies, Kyoto University, Yoshida-nihonmatsu-cho, Sakyo-ku, Kyoto 606-8316, Japan.

[2]Glass Mechanics, Physics Institute, UMR-CNRS 6251, Beaulieu campus, University of Rennes, 35042 Rennes cedex, France.

[3]CNRS, ISCR (Institut des Sciences Chimiques de Rennes) - UMR 6226, F 35000 Rennes, France.

## 1. First-principles calculations

All density-functional theory (*DFT*) calculations were performed with the Vienna Ab initio Simulation Package (*VASP*) *[Kresse & Furthmüller, 1996]* using the projector-augmented-wave (*PAW*) method *[Blöchl, 1994; Kresse & Joubert, 1999]* to describe the interaction between the valence electrons and the frozen ionic cores. The exchange–correlation energy was treated within the generalised-gradient approximation (*GGA*), using the *PBEsol* functional *[Perdew et al., 2008]*, a revision of the standard *PBE* parametrisation that is optimised for densely packed solids and their surfaces and is known to reproduce experimental lattice parameters and elastic constants of oxide and chalcogenide solids more accurately than *PBE*. *PBEsol* was selected through the INCAR tag `GGA = PS`.

In all calculations, plane waves were expanded up to a kinetic-energy cutoff of **ENCUT = 520 eV**, roughly 1.3 times the highest default cutoff of the PAW potentials used, in order to converge the stress tensor to the accuracy required for elastic-constant calculations. The precision flag was set to `PREC = Accurate`, real-space projection was disabled (`LREAL = .FALSE.`) for small cells and switched to `LREAL = Auto` only for the larger super-cells (more than ~20 atoms), where this is known not to alter the stresses. Self-consistent electronic iterations were stopped when the change in total energy between two successive cycles fell below **EDIFF = $10^{-8}$ eV** (with up to 100 electronic steps), using the `ALGO = Normal` (blocked-Davidson) electronic minimiser, which is more stable than the default fast scheme for the wide-gap insulators considered here. Partial occupancies were treated with Gaussian smearing (`ISMEAR = 0`) and a small electronic broadening of **σ = 0.05 eV**, appropriate for non-metallic systems. Spin polarisation was switched off (`ISPIN = 1`) since all the compounds investigated here are non-magnetic. The Brillouin zone was sampled with a Γ-centred Monkhorst–Pack grid *[Monkhorst & Pack, 1976]* with a converged density chosen for each compound.

For every compound, the experimental crystal structure was used as starting geometry. Atomic positions, cell shape and cell volume were then optimised simultaneously with `ISIF = 3` and `IBRION = 2` (conjugate-gradient ionic relaxation, up to `NSW = 200` steps) until the residual Hellmann–Feynman force on every ion fell below **|F| = $10^{-3}$ eV·Å$^{-1}$** (`EDIFFG = −10⁻³`) and the residual diagonal stresses dropped below ≈ 0.1 kbar. The relaxation was restarted from the previous CONTCAR until the cell parameters were stable to better than $10^{-4}$ Å between two successive runs. The relaxed geometries were then compared with the experimental cell parameters; the relative deviation $\Delta = (a_{theo} - a_{exp}) / a_{exp} \times 100$ % was computed for each lattice parameter, lattice angle and unit-cell volume in order to quantify the residual error of the functional.

## 2. Elastic stiffness tensor

The full second-order elastic stiffness tensor $C_{ij}$ was then evaluated for each fully relaxed structure with the finite-difference scheme implemented in *VASP [Le Page & Saxe, 2002]* by setting `IBRION = 6` together with `ISIF = 3`. In this approach, six independent finite Lagrangian distortions (three normal and three shear) are applied to the relaxed unit cell with both positive and negative amplitudes; for each distortion the internal atomic positions are re-optimised at fixed cell shape, the stress tensor is evaluated, and the elements of $C_{ij}$ are obtained by central finite differences of the stresses with respect to the strains. We used a strain amplitude of **POTIM = 0.015** and **NFREE = 4** displacements per independent strain (two positive and two negative steps), so that $C_{ij}$ is obtained from a four-point central-difference stencil and the leading $O(\varepsilon^2)$ error is suppressed. Because the ions are allowed to relax at every distorted geometry, the resulting $C_{ij}$ includes the ionic-relaxation contribution in addition to the purely clamped-ion electronic response and therefore corresponds directly to the static, low-frequency elastic constants accessible to experiment.

The elastic step was started from the converged charge density and wave function of the preceding relaxation (`ISTART = 1`, `ICHARG = 1`) and used the same plane-wave cutoff (520 eV), the same electronic convergence criterion ($10^{-8}$ eV), the same Gaussian smearing ($\sigma = 0.05$ eV) and the same k-point grid as the relaxation. To ensure numerical stability of the finite-difference stresses, the parallelization was set with *`NPAR`* equal to the total number of MPI ranks, as required by VASP for `IBRION = 6`, and the redistribution of k points across finite-difference steps was disabled with `KBLOWUP = .FALSE.` to avoid the spurious k-point reshuffling that can otherwise occur on lower-symmetry distortions.

Because *VASP* prints the elastic tensor in its native Voigt ordering *(1 = XX, 2 = YY, 3 = ZZ, 4 = XY, 5 = YZ, 6 = ZX)*, whereas the convention used by the *ELATE* analysis tool *[Gaillac, Pullumbi & Coudert, 2016]* and most of the post-Voigt literature is *(1 = XX, 2 = YY, 3 = ZZ, 4 = YZ, 5 = ZX, 6 = XY)*, the rows and columns 4–6 of the *VASP* output were systematically reordered before any subsequent analysis. Stiffness coefficients were also converted from kbar to GPa (factor 0.1).

## 3. Mechanical (Born) stability

The mechanical stability of each phase was tested using the generalized Born criterion, which requires the 6 × 6 stiffness matrix expressed in Voigt notation to be positive definite *[Born & Huang, 1954; Mouhat & Coudert, 2014]*. This is equivalent to demanding that all eigenvalues $\lambda_k$ of $C$ satisfy $\lambda_k > 0$; in practice, the eigenvalues of the symmetrized tensor were diagonalized numerically and a phase was accepted as elastically stable only when this condition was satisfied.

## 4. Polycrystalline elastic moduli

Although the materials investigated here are anisotropic single crystals, most experimentally accessible quantities (powder or ceramic samples, indentation moduli, sound velocities of sintered pellets, etc.) correspond to polycrystalline averages over randomly oriented grains. We therefore evaluated the bulk modulus $K$, shear modulus $G$, Young's modulus $E$ and Poisson's ratio $\nu$ using the standard Voigt *[Voigt, 1928]*, Reuss *[Reuss, 1929]* and Hill *[Hill, 1952]* averaging schemes. The Voigt (uniform-strain, upper-bound) bulk and shear moduli are obtained directly from the components of $C$:

$$9\,K_V = (C_{11} + C_{22} + C_{33}) + 2\,(C_{12} + C_{13} + C_{23}),$$

$$15\,G_V = (C_{11} + C_{22} + C_{33}) - (C_{12} + C_{13} + C_{23}) + 3\,(C_{44} + C_{55} + C_{66}),$$

while the Reuss (uniform-stress, lower-bound) averages are expressed through the compliance tensor $S = C^{-1}$:

$$1\,/\,K_R = (S_{11} + S_{22} + S_{33}) + 2\,(S_{12} + S_{13} + S_{23}),$$

$$15\,/\,G_R = 4\,(S_{11} + S_{22} + S_{33}) - 4\,(S_{12} + S_{13} + S_{23}) + 3\,(S_{44} + S_{55} + S_{66}).$$

The Hill average, generally taken as the best estimate for an isotropic polycrystal, is the arithmetic mean of the two bounds, $K_H = ½\,(K_V + K_R)$ and $G_H = ½\,(G_V + G_R)$. The corresponding Young's modulus and Poisson's ratio were derived from each ($K$, $G$) pair using the standard isotropic relations

$$E = \frac{9KG}{3K + G}$$

$$\nu = \frac{3K - 2G}{2(3K + G)}$$

The Pugh ratio $k = K_H / G_H$ *[Pugh, 1954]* was used as an empirical indicator of the ductile/brittle character (ductile for $k > 1.75$) and the related quantity $3\,K / G$, proportional to the ratio of resistance to volume change versus shape change, was also tabulated.

## 5. Automated extraction workflow

To handle the full set of compounds in a reproducible way, an in-house Python 3 script was developed to post-process the *VASP* output. For each calculation, the script (i) parses the lattice type, the number of space-group operations, the static point symmetry and the point group of the full space group as identified by *VASP* from the final *OUTCAR*; (ii) extracts the lattice vectors of the experimental cell from the input *POSCAR* and those of the relaxed cell from the last *"direct lattice vectors"* block of the *OUTCAR*, converts them to ($a$, $b$, $c$, $\alpha$, $\beta$, $\gamma$, $V$) and computes the relative deviation between the two; (iii) reads the *"TOTAL ELASTIC MODULI (kBar)"* block, reorders it from the *VASP* convention to the *ELATE* convention and converts it to GPa; (iv) computes the Voigt, Reuss and Hill averages of $K$, $G$, $E$ and $\nu$ together with the Pugh ratio and the related $3\,K / G$ indicator; and (v) tests the Born stability criterion by diagonalising $C$. The script writes a single, human-readable report per compound that contains all of these quantities, ensuring that the same definitions, the same Voigt convention and the same averaging formulas are applied uniformly to every structure analyzed in this work.

***First-principles calculations***

***Elastic stiffness tensor***

***Mechanical (Born) stability***

***Polycrystalline elastic moduli***

**Table S1**. Experimental and *PBEsol-DFT* cell parameters of the studied compounds in their *conventional* setting, with the relative deviation $\Delta = (X_{theo} - X_{exp}) / X_{exp}$ expressed in %. Lengths in Å, angles in °, volumes in Å³. *T* is the temperature of the experimental measurement; *ICSD* entries are given for cross-reference.

| Compound | Crystal system (*SG*) | Source | *a* (Å) | *b* (Å) | *c* (Å) | *α* (°) | *β* (°) | *γ* (°) | *V* (Å³) | *T* (K) | *ICSD* | Ref. |
|---|---|---|---|---|---|---|---|---|---|---|---|---|
| $SrAl_2O_4$ | Hexagonal ($P6_3$) | *Exp.* | 8.9291 | 8.9291 | 8.4963 | 90.00 | 90.00 | 120.00 | 586.65 | 983 | 160298 | **[1]** *Avdeev et al., 2007* |
| | | *PBEsol* | 8.9088 | 8.9088 | 8.3842 | 90.00 | 90.00 | 120.00 | 576.28 | | | |
| | | *Δ* (%) | -0.23 | -0.23 | -1.32 | 0.00 | 0.00 | 0.00 | -1.77 | | | |
| $SrAl_2O_4$ | Monoclinic ($P2_1$) | *Exp.* | 8.4436 | 8.8224 | 5.1596 | 90.00 | 93.41 | 90.00 | 383.68 | 298 | 160296 | **[1]** *Avdeev et al., 2007* |
| | | *PBEsol* | 8.4309 | 8.8016 | 5.1692 | 90.00 | 93.58 | 90.00 | 382.84 | | | |
| | | *Δ* (%) | -0.15 | -0.24 | +0.19 | 0.00 | +0.17 | 0.00 | -0.22 | | | |
| $BaAl_2O_4$ | Hexa. high T ($P6_322$) | *Exp.* | 5.227 | 5.227 | 8.802 | 90.00 | 90.00 | 120.00 | 208.27 | 293 | 21080 | **[2]** *Do Dinh Chieu & Bertaut, 1965* |
| | | *PBEsol* | 5.2011 | 5.2011 | 8.9026 | 90.00 | 90.00 | 120.00 | 208.56 | | | |
| | | *Δ* (%) | -0.50 | -0.50 | +1.14 | 0.00 | 0.00 | 0.00 | +0.14 | | | |
| $BaAl_2O_4$ | Hexa. low T ($P6_3$) | *Exp.* | 10.4508 | 10.4508 | 8.7907 | 90.00 | 90.00 | 120.00 | 831.48 | 293 | 209846 | **[3]** *Saines et al., 2006* |
| | | *PBEsol* | 10.4642 | 10.4642 | 8.7763 | 90.00 | 90.00 | 120.00 | 832.25 | | | |
| | | *Δ* (%) | +0.13 | +0.13 | -0.16 | 0.00 | 0.00 | 0.00 | +0.09 | | | |
| $Ba_2MgSi_2O_7$ | Tetragonal ($P\text{-}42_1m$) | *Exp.* | 8.2036 | 8.2036 | 5.4058 | 90.00 | 90.00 | 90.00 | 363.81 | 296 | 81117 | **[4]** *Shimizu et al., 1995* |
| | | *PBEsol* | 8.2197 | 8.2197 | 5.4017 | 90.00 | 90.00 | 90.00 | 364.96 | | | |
| | | *Δ* (%) | +0.20 | +0.20 | -0.08 | 0.00 | 0.00 | 0.00 | +0.32 | | | |
| $Ba_2MgSi_2O_7$ | Monoclinic ($C2/c$) | *Exp.* | 8.4171 | 10.7194 | 8.4501 | 90.00 | 110.77 | 90.00 | 712.87 | 293 | 183983 | **[5]** *Ardit et al., 2012* |
| | | *PBEsol* | 8.4240 | 10.7136 | 8.4621 | 90.00 | 110.92 | 90.00 | 713.38 | | | |
| | | *Δ* (%) | +0.08 | -0.05 | +0.14 | 0.00 | +0.15 | 0.00 | +0.07 | | | |
| $Ca_2MgSi_2O_7$ | Tetragonal ($P\text{-}42_1m$) | *Exp.* | 7.8338 | 7.8338 | 5.0088 | 90.00 | 90.00 | 90.00 | 307.38 | 298 | 192373 | **[6]** *Dondi et al., 2014* |
| | | *PBEsol* | 7.8195 | 7.8195 | 4.9764 | 90.00 | 90.00 | 90.00 | 304.28 | | | |
| | | *Δ* (%) | -0.18 | -0.18 | -0.65 | 0.00 | 0.00 | 0.00 | -1.01 | | | |
| $ZrO_2$ | Tetragonal ($P4_2/nmc$) | *Exp.* | 3.6030 | 3.6030 | 5.1739 | 90.00 | 90.00 | 90.00 | 67.17 | 293 | 123106 | **[7]** *Sirotinkin et al., 2022* |
| | | *PBEsol* | 3.5869 | 3.5869 | 5.1754 | 90.00 | 90.00 | 90.00 | 66.59 | | | |
| | | *Δ* (%) | -0.45 | -0.45 | +0.03 | 0.00 | 0.00 | 0.00 | -0.87 | | | |
| $ZrO_2$ | Monoclinic ($P2_1/c$) | *Exp.* | 5.1450 | 5.2075 | 5.3107 | 90.00 | 99.23 | 90.00 | 140.45 | 293 | 18190 | **[8]** *Smith & Newkirk, 1965* |
| | | *PBEsol* | 5.1302 | 5.2158 | 5.2958 | 90.00 | 99.57 | 90.00 | 139.73 | | | |
| | | *Δ* (%) | -0.29 | +0.16 | -0.28 | 0.00 | +0.34 | 0.00 | -0.51 | | | |
| ZnS | Cubic ($F\text{-}43m$) | *Exp.* | 5.3525 | 5.3525 | 5.3525 | 90.00 | 90.00 | 90.00 | 153.35 | 100 | 230706 | **[9]** *Thiodjio Sendja et al., 2018* |
| | | *PBEsol* | 5.3612 | 5.3612 | 5.3612 | 90.00 | 90.00 | 90.00 | 154.09 | | | |
| | | *Δ* (%) | +0.16 | +0.16 | +0.16 | 0.00 | 0.00 | 0.00 | +0.49 | | | |
| | | *Exp.* | 3.8227 | 3.8227 | 6.2607 | 90.00 | 90.00 | 120.00 | 79.23 | | | |
| ZnS | Hexagonal ($P6_3mc$) | *PBEsol* | 3.7856 | 3.7856 | 6.2134 | 90.00 | 90.00 | 120.00 | 77.11 | 293 | 67453 | **[10]** *Kisi & Elcombe, 1989* |
| | | *Δ* (%) | -0.97 | -0.97 | -0.76 | 0.00 | 0.00 | 0.00 | -2.68 | | | |
| | | *Exp.* | 3.9619 | 12.8541 | 6.1175 | 90.00 | 90.00 | 90.00 | 311.54 | | | |
| BaZnOS | Orthorhombic (*Cmcm*) | *PBEsol* | 3.9273 | 12.8310 | 6.0850 | 90.00 | 90.00 | 90.00 | 306.63 | 240 | 171239 | **[11]** *Broadley et al., 2005* |
| | | *Δ* (%) | -0.87 | -0.18 | -0.53 | 0.00 | 0.00 | 0.00 | -1.58 | | | |
| | | *Exp.* | 3.7573 | 3.7573 | 11.4013 | 90.00 | 90.00 | 120.00 | 139.39 | | | |
| CaZnOS | Hexagonal ($P6_3mc$) | *PBEsol* | 3.7307 | 3.7307 | 11.2759 | 90.00 | 90.00 | 120.00 | 135.91 | 293 | 245309 | **[12]** *Sambrook et al., 2007* |
| | | *Δ* (%) | -0.71 | -0.71 | -1.10 | 0.00 | 0.00 | 0.00 | -2.49 | | | |
| | | *Exp.* | 12.4770 | 4.6850 | 13.9440 | 90.00 | 93.54 | 90.00 | 813.54 | | | |
| $Ba_4Si_6O_{16}$ | Monoclinic ($P2_1/c$) | *PBEsol* | 12.5057 | 4.6858 | 13.9797 | 90.00 | 93.55 | 90.00 | 817.63 | 293 | 100310 | **[13]** *Hesse & Liebau, 1980* |
| | | *Δ* (%) | +0.23 | +0.02 | +0.26 | 0.00 | +0.01 | 0.00 | +0.50 | | | |
| | | *Exp.* | 32.6750 | 4.6950 | 13.8940 | 90.00 | 98.10 | 90.00 | 2110.20 | | | |
| $Ba_5Si_8O_{21}$ | Monoclinic (*C2/c*) | *PBEsol* | 32.9351 | 4.6972 | 13.9172 | 90.00 | 98.14 | 90.00 | 2131.32 | 293 | 100311 | **[13]** *Hesse & Liebau, 1980* |
| | | *Δ* (%) | +0.80 | +0.05 | +0.17 | 0.00 | +0.04 | 0.00 | +1.00 | | | |

**Notes:** all values correspond to the conventional unit cell. For phases whose *VASP* calculation was performed in the primitive setting (cubic *F*-43*m*, base-centred *C*mcm and *C*2/*c*), the relaxed primitive cell was converted analytically to the conventional cell prior to comparison. Volumes for monoclinic cells are computed as $V = abc \sin \beta$. *T* (K) is the measurement temperature reported in the original publication; *ICSD* entries refer to the Inorganic Crystal Structure Database collection codes.

## References (experimental cells)